\documentclass[twocolumn,aps,pra,showpacs,superscriptaddress]{revtex4}

\usepackage{graphicx}
\usepackage{amsmath,amssymb}
\usepackage{color}
\usepackage{epsfig}

\def\endproof{\vrule height6pt width6pt depth0pt}

\begin{document}


\title{Basic exclusivity graphs in quantum correlations}



\author{Ad\'an Cabello}
 \email{adan@us.es}
 \affiliation{Departamento de F\'{\i}sica Aplicada II, Universidad de
 Sevilla, E-41012 Sevilla, Spain}
\author{Lars Eirik Danielsen}
 \affiliation{Department of Informatics,
 University of Bergen, P.O. Box 7803, N-5020 Bergen, Norway}
\author{Antonio J. L\'{o}pez-Tarrida}
 \affiliation{Departamento de F\'{\i}sica Aplicada II,
 Universidad de Sevilla, E-41012 Sevilla, Spain}
\author{Jos\'{e} R. Portillo}
 \affiliation{Departamento de Matem\'{a}tica Aplicada I,
 Universidad de Sevilla, E-41012 Sevilla, Spain}


\date{\today}



\begin{abstract}
A fundamental problem is to understand why quantum theory only violates some noncontextuality (NC) inequalities and identify the physical principles that prevent higher-than-quantum violations. We prove that quantum theory only violates those NC inequalities whose exclusivity graphs contain, as induced subgraphs, odd cycles of length five or more, and/or their complements. In addition, we show that odd cycles are the exclusivity graphs of a well-known family of NC inequalities and that there is also a family of NC inequalities whose exclusivity graphs are the complements of odd cycles. We characterize the maximum noncontextual and quantum values of these inequalities, and provide evidence supporting the conjecture that the maximum quantum violation of these inequalities is exactly singled out by the exclusivity principle.
\end{abstract}



\maketitle


\section{Introduction}
\label{Sec0}


Quantum contextuality, namely, the fact that the quantum correlations between the results of compatible measurements cannot be reproduced with noncontextual hidden variable (NCHV) theories \cite{Specker60,Bell66,KS67} is behind a wide spectrum of applications of quantum theory (QT) to communication and computation \cite{Ekert91,SBKTP09,AB09,BCMW10,NDSC12,CDNS12}. Quantum contextual correlations are experimentally detected through the violation of inequalities satisfied by NCHV models, called noncontextuality (NC) inequalities \cite{CFHR08, KCBS08, Cabello08, BBCP09}. A fundamental problem is to understand why QT only violates some NC inequalities and identify the physical principles that prevent higher-than-quantum violations of these inequalities \cite{PPKSWZ09,NW09,OW10,Cabello12b}.

In this paper we investigate this problem. In Sec.~\ref{Sec1} we introduce a tool that we will use throughout the paper, namely, the exclusivity graph. In Sec.~\ref{Sec2}, we present a necessary condition for the existence of quantum contextual correlations: We prove that QT violates only those NC inequalities whose exclusivity graphs contain, as induced subgraphs, odd cycles on five or more vertices and/or their complements. In Sec.~\ref{Sec3}, we show that a lower bound of the dimension (i.e., of the number of perfectly distinguishable states) of the quantum system that is used to violate an NC inequality can be obtained by identifying induced subgraphs in the exclusivity graph of the NC inequality.

The result in Sec.~\ref{Sec2} suggests that NC inequalities whose exclusivity graph is either an odd cycle or its complement are especially important for understanding the way QT violates NC inequalities. In Sec.~\ref{Sec4}, we show that each of these types of exclusivity graphs is connected to a family of NC inequalities and provide the quantum states and measurements leading to the maximum quantum violation. Finally, in Sec.~\ref{Sec5} we present some results that suggest that the exclusivity principle, namely, that the sum of the probabilities of a set of pairwise exclusive events cannot exceed~1, explains the maximum quantum violation of all the NC inequalities discussed in Sec.~\ref{Sec4}.


\section{The exclusivity graph of an NC inequality}
\label{Sec1}


\begin{figure}[tb]
\begin{center}
\centerline{\includegraphics[scale=0.36]{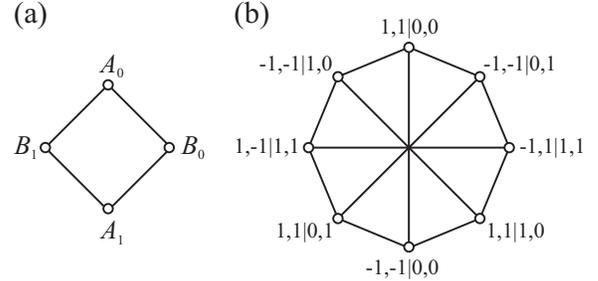}}
\caption{\label{Fig0}(a) The compatibility graph of the CHSH experiment. (b) The exclusivity graph of the CHSH inequality (\ref{CHSH2}).}
\end{center}
\end{figure}


Two different graphs can be associated to any given NC inequality. On one hand, the graph in which vertices represent the observables measured in the NC inequality and adjacent vertices represent those which are compatible \cite{AQBTC12}. This graph is the so-called compatibility graph. For example, consider the Clauser-Horne-Shimony-Holt (CHSH) inequality \cite{CHSH69},
\begin{equation}
 \beta = \langle A_0 B_0 \rangle + \langle A_0 B_1 \rangle + \langle A_1 B_0 \rangle - \langle A_1 B_1 \rangle \stackrel{\mbox{\tiny{ NCHV}}}{\leq} 2,
 \label{CHSH}
\end{equation}
where $\langle A_i B_j \rangle$ denotes the mean value of the product of the results of measuring the observables $A_i$ and $B_j$, each of them with possible results either $-1$ or $1$, and $\stackrel{\mbox{\tiny{ NCHV}}}{\leq} 2$ indicates that $2$ is the maximum value for $\beta$ for NCHV theories. In this inequality there are four observables: $A_0$, $B_0$, $A_1$, and $B_1$. All possible pairs of them are compatible except for the pairs $(A_0, A_1)$ and $(B_0, B_1)$. Therefore, the compatibility graph is the one depicted in Fig.~\ref{Fig0}(a).

On the other hand, by taking into account that
\begin{equation}
 \pm \langle A_i B_j \rangle = 2 [P( 1, \pm 1 \,|\, i, j)+P(-1, \mp 1 \,|\, i, j)]-1,
\end{equation}
where $P(a, b \,|\, i, j)$ is the probability of the event ``the result $a$ has been obtained when measuring $A_i$ and the result $b$ has been obtained when measuring $B_j$,'' inequality (\ref{CHSH}) can be written as
\begin{widetext}
\begin{equation}
\begin{split}
 S=\frac{\beta}{2}+2 &= P(1,1\,|\,0,0)+P(-1,-1\,|\,0,0)+P(1,1\,|\,0,1)+P(-1,-1\,|\,0,1) \\
 &+P(1,1\,|\,1,0)+P(-1,-1\,|\,1,0)+P(1,-1\,|\,1,1)+P(-1,1\,|\,1,1) \stackrel{\mbox{\tiny{ NCHV}}}{\leq} 3,
 \label{CHSH2}
\end{split}
\end{equation}
\end{widetext}
where the left-hand side is now a convex sum of probabilities of events. A new graph can be associated to the set of events, one in which the vertices represent the events and adjacent vertices represent events that cannot occur simultaneously (i.e., exclusive events). This is the so-called exclusivity graph \cite{CSW10}. For example, the exclusivity graph for the CHSH inequality (\ref{CHSH2}) is depicted in Fig.~\ref{Fig0}(b).

The interest of the exclusivity graph $G$ is that the maximum value of $S$ for NCHV theories is exactly given by the independence number of the graph $\alpha(G)$ (which is the maximum number of pairwise nonadjacent vertices in $G$), while the maximum value in QT is upper bounded (and frequently exactly given) by the Lov\'asz number of the graph $\vartheta(G)$ \cite{CSW10}.

The important point is that {\em any} experiment producing quantum contextual correlations can be associated to an exclusivity graph $G$ for which $\alpha(G) < \vartheta(G)$. Hereafter, we will refer to a graph with this property as a quantum contextual graph (QCG), and to a graph for which $\alpha(G)=\vartheta(G)$ as a quantum noncontextual graph (QNCG).

A subgraph $H$ of a graph $G$ is said to be induced if, for any pair of vertices $i$ and $j$ of $H$, $ij$ is an edge of $H$ if and only if $ij$ is an edge of $G$. For example, a graph $G$ has an induced pentagon if it is possible to remove from $G$ all but five vertices (and their corresponding edges) so that we end up with a pentagon with no additional edges.


\section{Basic exclusivity graphs}
\label{Sec2}


{\em Result 1.} The exclusivity graph of any NC inequality violated by QT contains, as induced subgraphs, odd cycles on five or more vertices and/or their complements.

\begin{figure}[t]
\begin{center}
\centerline{\includegraphics[scale=0.41]{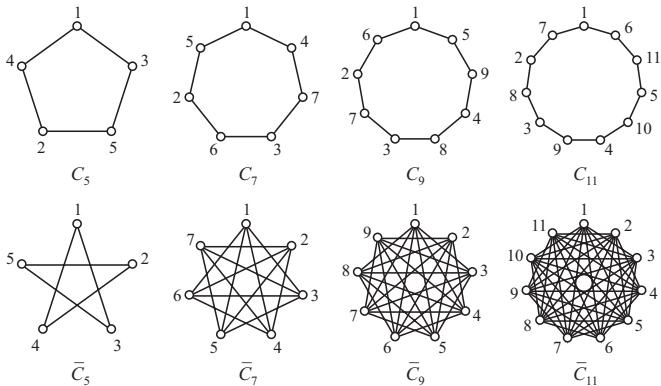}}
\caption{\label{Fig1}$C_j$ denote the odd cycles on $j$ vertices and $\bar{C}_j$ their complements. The figure depicts the cases $j=5,7,9,11$. Notice that $C_5$ and $\bar{C}_5$ are isomorphic.}
\end{center}
\end{figure}

This result is based on two fundamental results in graph theory. The strong perfect graph theorem and the (weak) perfect graph theorem.
Perfect graphs were introduced \cite{Berge61} in connection to the problem of the zero-error capacity of a noisy channel \cite{Shannon56}:
Shannon observed that $\omega(G^{\ast n})=\omega(G)^n$ for graphs such that $\omega(G)=\chi(G)$, which made the problem of characterizing the Shannon capacity of such graphs more tractable [$G^{\ast n}$ is the disjunctive product of $n$ copies of $G$, $\omega(G)$ is the clique number, and $\chi(G)$ is the chromatic number of $G$; in general, $\omega(G) \leq \chi(G)$]. Berge defined perfect graphs as those graphs $G$ for which $\omega(H)=\chi(H)$ for each induced subgraph $H \subseteq G$. Berge observed that all odd cycles $C_{n}$ with $n \ge 5$ (known in graph theory as ``odd holes'') and their complements $\bar{C}_{n}$ (known as ``odd antiholes'') satisfy $\omega(G)<\chi(G)$. From this result, Berge conjectured that a graph $G$ is perfect if and only if $G$ has no odd hole or odd antihole as induced subgraph (strong perfect graph conjecture). This conjecture has been recently proven, and it is now known as the strong perfect graph theorem \cite{CRST06}. The simplest odd holes and antiholes are illustrated in Fig.~\ref{Fig1}.

The perfect graph conjecture (due also to Berge), which was later proved by Lov\'asz \cite{Lovasz72} and is now known as the (weak) perfect graph theorem, states that if $G$ is a perfect graph, then its complement, $\bar{G}$, is also a perfect graph.


{\em Theorem 1.} Let $G$ be the exclusivity graph of an NC inequality. If $G$ is a perfect graph, then $G$ is a QNCG and, as a consequence, the NC inequality is not violated by QT.


{\em Proof.} If $G$ is perfect, then $\omega(G)=\chi(G)$. On the other hand, according to the sandwich theorem \cite{Knuth94}, $\omega(G) \le \vartheta(\bar{G}) \le \chi(G)$ for any graph $G$. Hence, $G$ perfect implies $\omega(G)=\vartheta(\bar{G})=\chi(G)$. Given that $\omega(G)=\alpha(\bar{G})$, we obtain $\alpha(\bar{G})=\vartheta(\bar{G})$, i.e., if $G$ is a perfect graph, then $\bar{G}$ is a QNCG. Applying now the (weak) perfect graph theorem, if $G$ is perfect, then $\bar{G}$ is also perfect. Therefore, repeating the previous argument the other way around we conclude that $G$ is a QNCG as well: $\alpha(G)=\vartheta(G)$. This finishes the proof. \hfill \endproof


{\em Corollary 1.} If $G$ is a QCG, i.e., if $G$ is the exclusivity graph of an NC inequality violated by QT, then $G$ is {\em not} perfect. Consequently, by the strong perfect graph theorem, $G$ must contain odd holes and/or odd antiholes as induced subgraphs.


This proves Result~1.


{\em Observation 1.} A NC inequality not violated by QT may contain an NC inequality violated by QT. This means that, even if the exclusivity graph of the initial NC inequality is not a QCG, the exclusivity graph of the initial NC inequality may have an induced subgraph that is a QCG.


{\em Observation 2.} No odd cycle or complement of an odd cycle has an odd cycle or a complement of an odd cycle as an induced subgraph. This suggests that odd cycles and their complements could be used as a basis for a decomposition of any exclusivity graph.


{\em Observation 3.} The basic exclusivity graphs necessary for quantum contextuality are the same basic graphs necessary for graph nonperfectness.


{\em Proof.} A graph $G$ is called almost perfect \cite{Padberg74,Padberg76} if there exists at least one vertex $v \in V(G)$ such that the graph $G-v$ obtained by deleting $v$ (and its incident edges) is perfect. Trivially, every perfect graph is almost perfect. There are many nonperfect graphs which are almost perfect. The interesting point is that there are nonperfect graphs for which the deletion of {\em any} vertex gives rise to a perfect graph: A graph $G$ is minimal (or minimally) imperfect if it is not perfect but {\em every} induced subgraph $H \subset G$ is perfect [i.e., $\omega(G)<\chi(G)$ but $\omega(H)=\chi(H), \forall H \subset G$, $H$ induced]. Note that, as a consequence of the strong perfect graph theorem, any nonperfect graph is either an odd hole or an odd antihole, or must contain such odd holes and/or odd antiholes as induced subgraphs. Hence, the {\em only} minimal imperfect graphs are the odd holes and odd antiholes. Moreover, given that every induced subgraph $H \subset G_m$ of a minimal imperfect graph $G_m$ is perfect, then the deletion of an arbitrary vertex $v \in V(G_m)$ produces an induced subgraph of $G_m$ which is a perfect graph \cite{PS01}. Therefore, minimal imperfect graphs are not only almost perfect but the only almost-perfect graphs for which almost perfectness does not depend on the vertex choice. That is why odd holes and antiholes are the necessary basic graphs for nonperfectness: Starting from an arbitrary nonperfect graph, and deleting vertex by vertex trying to preserve nonperfectness in the subsequent resulting induced subgraphs, eventually leads to a minimal imperfect graph. Since minimal imperfect graphs are also QCGs and all QCGs are nonperfect (Corollary~1), then minimal imperfect graphs are the simplest ones giving rise to quantum contextuality.\hfill \endproof


\section{Basic exclusivity graphs and the dimension of the quantum system}
\label{Sec3}


A distinguishing feature of the complements of the odd cycles is that their presence as induced subgraphs in an exclusivity graph $G$ provides valuable information about the minimum dimension that a quantum system must have in order to generate events reproducing all the relationships of exclusivity described by $G$. While any of the relationships of exclusivity in an odd cycle can be reproduced using a quantum three-dimensional system, this is impossible for the relationships of exclusivity described by the complement of an odd cycle with $n \ge 7$ vertices.


{\em Result 2.} If an exclusivity graph corresponding to an NC inequality has as an induced subgraph a complement of an odd cycle on $n \ge 5$ vertices, then the quantum dimension of the systems whose events reproduce these exclusivity relationships is, at least, $\lfloor 2n/3 \rfloor$.


{\em Proof: }A (faithful) orthonormal representation (OR) of $G$ is an assignment of unit vectors $\{|v_j\rangle\}_{j=1}^{n}$ to the vertices of $G$ such that orthogonal vectors are assigned to the vertices (iff) if they are adjacent. Vectors $\{|v_j\rangle\}_{j=1}^{n}$ represent the states of the system after the corresponding events; vectors are orthogonal if (iff) events are exclusive.

Let $G$ be an exclusivity graph containing an odd antihole $\bar{C_n}$ as induced subgraph. To obtain a lower bound of the dimension~$d$ of the quantum system producing events whose exclusivity relationships are described exactly by $G$, we can study the constraints imposed by the presence of $\bar{C_n}$. Note that for any two different vertices $u,v \in \bar{C_n}$, $N(u) \neq N(v)$, where $N(i)$ denotes the neighborhood of vertex $i$ (see Fig.~\ref{Fig1}). This implies that a faithful OR of $G$ must assign different vectors to $u$ and $v$. As a consequence, we can lower bound $d$ by identifying subgraphs in $\bar{C_n}$ which are geometrically impossible in a space of lower dimension, assuming that distinct vertices are assigned distinct vectors. For example, the simplest impossible graph in $d=1$ consists of two nonadjacent vertices in $\bar{C_n}$; in $d=2$, three vertices, one of them adjacent to the other two. From these two impossible graphs, one can recursively construct impossible graphs in any $d$ by adding two vertices adjacent to all vertices of an impossible graph in $d-2$. For example, if $\bar{C_n}$ contains a square, then $d > 3$. In brief, the graph $F$ obtained by deleting $\lfloor d/2 \rfloor$ disjoint edges from a $d$-vertex complete graph $K_d$ is an impossible subgraph in dimension~$d-1$ of $\bar{C_n}$.

To lower bound $d$, we must consider three cases: (C1) If $n=3m, m \in \mathbb{N}$, take the subgraph $F$ induced in $\bar{C_n}$ by vertices $\{1, 2, 4, 5, \ldots, 3i+1, 3i+2, \ldots, 3(m-1)+1, 3(m-1)+2 \}$ (see Fig.~\ref{Fig1}). $F$ is isomorphic to $K_{2m}$ minus $m$ disjoint edges and is, therefore, an impossible graph in $d=2m-1$. Hence, $\bar{C_n}$ is not faithfully representable in $d=2m-1=\frac{2n}{3}-1=\lfloor 2n/3 \rfloor$-1. The same holds true for $G\supseteq \bar{C_n}$. (C2) If $n=3m+1, m \in \mathbb{N}$, take the subgraph $F$ induced in $\bar{C_n}$ by vertices $\{1, 2, 4, 5,\ldots, 3i+1, 3i+2,\ldots, 3(m-1)+1, 3(m-1)+2 \}$. $F$ is isomorphic to $K_{2m}$ minus $m$ disjoint edges and is, therefore, an impossible graph in $d=2m-1$. Hence, $\bar{C_n}$ is not faithfully representable in $d=2m-1=\frac{2n-1}{3}-1=\lfloor 2n/3 \rfloor$-1. The same holds true for $G\supseteq \bar{C_n}$. (C3) If $n=3m+2, m \in \mathbb{N}$, take the subgraph $F$ induced in $\bar{C_n}$ by vertices $\{1, 2, 4, 5,\ldots, 3i+1, 3i+2, \ldots, 3(m-1)+1, 3(m-1)+2, 3m+1 \}$. $F$ is isomorphic to $K_{2m+1}$ minus $m$ disjoint edges and, therefore, it is an impossible graph in $d=2m$. Hence, $\bar{C_n}$ is not faithfully representable in $d=2m=\frac{2(n-2)}{3}=\lfloor 2n/3 \rfloor$-1. The same holds true for $G\supseteq \bar{C_n}$.\hfill \endproof


\section{NC inequalities represented by basic exclusivity graphs}
\label{Sec4}


{\em Result 3.} For any cycle $C_{n}$ with $n$ odd $\ge 5$, there is an NC inequality such that
\begin{equation}
 S(C_n) \stackrel{\mbox{\tiny{ NCHV}}}{\leq} \alpha(C_{n}) \stackrel{\mbox{\tiny{Q}}}{\leq} \vartheta(C_{n}),
 \label{result3}
\end{equation}
where $S(C_n)$ is a sum of probabilities of events matching the relationships of exclusivity represented by $C_n$, $\stackrel{\mbox{\tiny{Q}}}{\leq} \vartheta(C_{n})$ indicates that its maximum value in QT is {\em exactly} $\vartheta(C_{n})$, and
\begin{subequations}
\begin{align}
 \alpha(C_{n})&=\frac{n-1}{2},\\
 \vartheta(C_{n})&=\frac{n\cos\left(\frac{\pi}{n}\right)}{1+\cos\left(\frac{\pi}{n}\right)}
\end{align}
\end{subequations}
are, respectively, the independence number and the Lov\'asz number of $C_{n}$.


{\em Proof.} By explicit construction. For any $n$ odd $\ge 5$, the events in the following sum of probabilities of events:
\begin{equation}
 S(C_n)=\sum_{i=1}^n P(1,0|i,i+[\frac{n}{2}]),
\end{equation}
where the sum in each event is taken modulo $n$, and numerating the vertices of $C_n$ as in Fig.~\ref{Fig1}, have exactly the relationships of exclusivity represented by $C_n$.

The fact that the maximum value of $S(C_n)$ for NCHV theories is $\alpha(C_n)$ is proven in \cite{CSW10}.
The fact that $\vartheta(C_n)$ is not only an upper bound of the maximum quantum value, but a value that QT actually reaches can be seen by preparing the system in the quantum state
\begin{equation}
 \langle \psi |=(1,0,0) \label{state1}
\end{equation}
and measuring the observables represented by
\begin{equation}
 j = | v_j \rangle \langle v_j |,
 \label{observable}
\end{equation}
where
\begin{equation}
\langle v_j |=\left[\cos \phi, \sin \phi \cos \left(\frac{2 \pi j}{n}\right),\sin \phi \sin \left(\frac{2 \pi j}{n}\right)\right], \label{vectors1}
\end{equation}
with $j=1,\ldots,n$, $\cos^2 \phi =\frac{\vartheta(C_{n})}{n}$.\hfill \endproof

The vectors (\ref{vectors1}) constitute a Lov\'asz-optimum OR of $C_{n}$. An OR is Lov\'asz optimum if there is a unit vector $|\psi\rangle$, called handle, such that $\sum_{j=1}^{n} |\langle v_j | \psi \rangle |^2 = \vartheta(G)$. In our case, the handle is given by Eq.~(\ref{state1}).


{\em Result 4.} For any complement of a cycle $\bar{C}_{n}$ with $n$ odd $\ge 5$, there is an NC inequality such that
\begin{equation}
 S(\bar{C}_{n}) \stackrel{\mbox{\tiny{ NCHV}}}{\leq} \alpha(\bar{C}_{n}) \stackrel{\mbox{\tiny{Q}}}{\leq} \vartheta(\bar{C}_{n}),
 \label{result4}
\end{equation}
where
\begin{subequations}
\begin{align}
 \alpha(\bar{C}_{n})&=2,\\
 \vartheta(\bar{C}_{n})&=\frac{1+\cos\left(\frac{\pi}{n}\right)}{\cos\left(\frac{\pi}{n}\right)}.
\end{align}
\end{subequations}


{\em Proof.} For $n=5$, the proof of Result~3 is valid, since $C_5$ and $\bar{C}_5$ are isomorphic. For any $n$ odd $\ge 7$, the events in the following sum of probabilities of events:
\begin{equation}
 S(\bar{C}_n)=\sum_{i=1}^n P(1,0,\ldots,0|i,i+2,\ldots,i+n-3),
\end{equation}
where the sum in each event is taken modulo $n$, and numerating the vertices of $\bar{C}_n$ as in Fig.~\ref{Fig1}, have all the relationships of exclusivity represented by $\bar{C}_n$.

The fact that the maximum value of $S(\bar{C}_n)$ for NCHV theories is $\alpha(\bar{C}_n)$ is proven in \cite{CSW10}.
The fact that $\vartheta(\bar{C}_n)$ is not only an upper bound of the maximum quantum value, but a value that QT actually reaches can be seen by preparing the system in the quantum state
\begin{equation}
 \langle \psi|=(1,0,\ldots,0), \label{state2}
\end{equation}
and measuring the observables represented by
\begin{equation}
 j = | v_j \rangle \langle v_j |,
\end{equation}
where the $k$-th component of $\langle v_j|$, denoted as $v_{j,k}$, with $0 \le j \le n-1$ and $0 \le k \le n-3$, is given by
\begin{subequations}
 \label{vectors2a}
\begin{align}
v_{j,0}&=\sqrt{\frac{\vartheta(\bar{C_n})}{n}}, \\
v_{j,2m-1}&= T_{j,m} \cos{(R_{j,m})}, \\
v_{j,2m}&= T_{j,m} \sin{(R_{j,m})},
\end{align}
\end{subequations}
where $m=1,2,\ldots, \frac{n-3}{2}$ and
\begin{subequations}
 \label{vectors2b}
\begin{align}
T_{j,m}&= (-1)^{j(m+1)}
\sqrt{\frac{2 \left(\cos{(\frac{\pi}{n})}+(-1)^{m+1}\cos{\left[\frac{(m+1) \pi}{n}\right]}\right)}{n \cos(\frac{\pi}{n})}},\\
R_{j,m}&=\frac{j (m+1)\pi}{n}.
\end{align}
\end{subequations}
\hfill \endproof

The vectors defined by Eqs.~(\ref{vectors2a}) and (\ref{vectors2b}) constitute a Lov\'asz-optimum OR of $\bar{C}_n$ with handle given by Eq.~(\ref{state2}).

Note that, for every NC inequality presented in this section, the compatibility graph of the observables is isomorphic to the exclusivity graph of the events. This follows from the fact that every observable in these NC inequalities is of the form~(\ref{observable}) and every event is of the type $1,0,\ldots,0|i,j,\ldots,z$.


\section{The exclusivity principle explains the quantum violation of the NC inequalities represented by basic exclusivity graphs}
\label{Sec5}


It has been recently proved that the principle that the sum of probabilities of a set of pairwise exclusive events cannot be higher than~1 \cite{Specker60,CSW10,Wright78,Specker09,Cabello12b,FSABCLA12,Henson12,Cabello12c}, which we will hereafter call the exclusivity (E) principle, exactly singles out the maximum quantum value for the NC inequality associated to $C_5$ \cite{Cabello12b}. The principle of local orthogonality \cite{FSABCLA12} may be seen as the E principle restricted to Bell scenarios \cite{Henson12}. Note, however, that while for a given graph $G$, there is always an NC inequality for which QT reaches $\vartheta(G)$ \cite{CSW10}, this is not true if ``NC inequality'' is replaced by ``Bell inequality'' \cite{SBBC13}.

A natural question, especially important in the light of Result~1, is whether the E principle may single out the maximum quantum value for the NC inequalities associated to $C_{n}$ and $\bar{C}_{n}$ for $n$ odd $\ge 7$ presented in Sec.~\ref{Sec4}.


{\em Observation 4.} ``[T]he evidence that the Shannon capacity of odd cycles is extremely close to the value of the Lov\'asz theta function'' \cite{CGR03} strongly suggests that the E principle singles out the maximum quantum value for the NC inequalities (\ref{result4}). This implies that the E principle also singles out the maximum quantum value for the NC inequalities (\ref{result3}). In any case, it is very unlikely that any actual experiment would allow us to distinguish nature's maximum violation (assumed to be given by QT) from the maximum value allowed by the E principle.


{\em Proof.} Reference~\cite{CSW10} shows that, for a given exclusivity graph $G$, the maximum value satisfying the E principle (applied only to one copy of $G$) for the sum $S (G)$ of probabilities of events whose relationships of exclusivity are represented by $G$ is given by the fractional packing number of $G$, $\alpha^* (G,\Gamma)$, which is equal to $\max \sum_{i\in V(G)} w_i$, where the maximum is taken over all $0 \leq w_i\leq 1$ and for all cliques (subsets of pairwise linked vertices) $C_j \in \Gamma$, under the restriction $\sum_{i \in C_j} w_i \leq 1$ \cite{Shannon56,SU97}. When $\Gamma$ is the set of all cliques of $G$, then $\alpha^* (G,\Gamma)$ is also called the Rosenfeld number $p(G)$ \cite{Rosenfeld67}. As shown in \cite{Cabello12b}, for vertex transitive graphs (such us, e.g., $C_n$ and $\bar{C}_n$), the maximum value of $S(G)$ satisfying the E principle applied to $n$ copies is given by $\sqrt[n]{p(G^{\ast n})}$, where $G$ is the exclusivity graph of the NC inequality. Recall that the maximum quantum value is given by $\vartheta(G)$. For $C_5$, $p(C_{5})=\frac{5}{2}$ and $\sqrt{p(C_{5} \ast C_{5})}=\vartheta(C_5)=\sqrt{5}$, which proves that the E principle applied to two copies of $C_5$ singles out the maximum quantum value of $S(\bar{C}_{5})$. The whole process can be summarized in the following expression:
\begin{equation}
 S(\bar{C}_{5}) \stackrel{\mbox{\tiny{ NCHV}}}{\leq} 2 \stackrel{\mbox{\tiny{ Q, E2}}}{\leq} \sqrt{5} \stackrel{\mbox{\tiny{ E1}}}{\leq} \frac{5}{2},
\end{equation}
where $\stackrel{\mbox{\tiny{ NCHV}}}{\leq} 2$ indicates that the maximum value for NCHV theories is $2$, $\stackrel{\mbox{\tiny{ Q, E2}}}{\leq} \sqrt{5}$ indicates that the maximum value for QT and for theories satisfying the E principle applied to two copies of $C_5$ is $\sqrt{5}$, and $\stackrel{\mbox{\tiny{ E1}}}{\leq} \frac{5}{2}$ indicates that the maximum value for theories satisfying the E principle applied to one copy of $C_5$ is $\frac{5}{2}$.

For $C_7$, with $n=1,2,3$, $\sqrt[n]{p(C_{7}^{\ast n})}=\frac{7}{2}$. Since $p(C_{7})=\frac{7}{2}$, this means that, for $C_7$, the E principle applied to two or three copies of $C_7$ does not tell us more than the E principle applied to one copy. However, the E principle applied to four copies of $C_7$ leads us to a value closer to the maximum quantum one. This follows from \cite{BB03}, where it is proven that
\begin{equation}
 \sqrt[4]{p(C_{7}^{\ast 4})} \le \frac{7}{\sqrt[4]{17}}\approx 3.4474.
\end{equation}
The situation is similar for $C_n$ with $n$ odd $\ge 9$: the value singled out by the E principle seems to converge to the maximum quantum one as more copies of $C_n$ are taken into account, but there is no clear evidence that this actually happens.

However, note that odd cycles are sparser than their complements. As a result, when the E principle is applied to multiple copies of complements of odd cycles, the resulting value approaches the maximum quantum one much faster. Consider, for example, $\bar{C}_7$. From \cite{VZ98}, we obtain that
\begin{equation}
 p(\bar{C}_{7}) = \frac{7}{3} > \sqrt{p(\bar{C}_{7}^{\ast 2})}=\frac{7}{\sqrt{10}} > \sqrt[3]{p(\bar{C}_{7}^{\ast 3})}=\frac{7}{\sqrt[3]{33}}.
\end{equation}
Therefore,
\begin{equation}
 S(\bar{C}_{7}) \stackrel{\mbox{\tiny{ NCHV}}}{\leq} 2 \stackrel{\mbox{\tiny{Q}}}{\leq} 2.1099 \stackrel{\mbox{\tiny{ E3}}}{\leq} 2.1824 \stackrel{\mbox{\tiny{ E2}}}{\leq} 2.2136 \stackrel{\mbox{\tiny{ E1}}}{\leq} 2.3333.
\end{equation}
For four copies of $\bar{C}_7$, only bounds of $\omega(\bar{C}_7^{\ast 4})$ are known \cite{VZ02}, but it is clear that the value is even closer to the maximum quantum one after considering the fourth copy, since
\begin{equation}
 \frac{7}{\sqrt[4]{115}}\approx 2.1376 \le \sqrt[4]{p(\bar{C}_{7}^{\ast 4})}\le \frac{7}{\sqrt[4]{108}} \approx 2.1714.
\end{equation}
For five copies, the best bounds known \cite{VZ02} are also compatible with a value even closer to the maximum quantum one.

The point is that there is strong evidence \cite{CGR03} that the Shannon capacity of odd cycles, $\Theta(C_n)$, is extremely close to their Lov\'asz number, $\vartheta(C_n)$. This is important since
\begin{equation}
 \lim_{m \rightarrow \infty} \sqrt[m]{p(\bar{C}_n^{\ast m})}= \frac{n}{\lim_{m \rightarrow \infty} \sqrt[m]{\omega(\bar{C}_n^{\ast m})}}=\frac{n}{\Theta(C_n)}
\end{equation}
(this holds because $C_n$ and $\bar{C}_n$ are vertex transitive), which, if $\Theta(C_n)=\vartheta(C_n)$, would be equal to $\vartheta(\bar{C}_n)$, which is exactly the maximum quantum value for $\bar{C}_n$.

If the E principle singles out the maximum quantum value of inequalities (\ref{result4}), then it also singles out the maximum quantum value of inequalities (\ref{result3}). To see it, consider one experiment testing the inequality (\ref{result4}) for a given $n$, and a completely independent experiment testing the inequality (\ref{result3}) for the same $n$. Notice that the exclusivity graph of the events defined by taking one event of the first inequality and the corresponding event of the second inequality is the complete graph on $n$ vertices, since the exclusivity graphs of the two inequalities are complementary. Therefore, the E principle imposes that the sum of the probabilities of the $n$ events constructed this way cannot exceed~$1$. Take into account that each of these probabilities is the product of the probability of the event in the first inequality times the probability of the corresponding event in the
second inequality, since the corresponding experiments are independent. Assume now that the E principle predicts that the maximum of the first inequality is $\vartheta(\bar{C_n})$ and that the maximum value of the first inequality is reached when all the probabilities are equal. Assuming that the maximum value of the second inequality is also reached when all the probabilities are equal, and recalling that $\vartheta(C_n) \vartheta(\bar{C_n})=n$, we are led to the conclusion that the maximum value allowed by the E principle for the second inequality cannot be higher than $\vartheta(C_n)$, which is precisely the maximum value in QT.\hfill \endproof


\section{Conclusions}
\label{Sec6}


Here we have proven that QT only violates a particular kind of NC inequalities: those whose exclusivity graph contains some basic subgraphs, odd cycles with five or more vertices and their complements (Result~1).
We have also shown that the presence of some of these subgraphs provides a lower bound to the minimum dimension that a quantum system must have to make the corresponding NC inequality experimentally testable (Result~2).

In addition, we have shown that there is a family of NC inequalities violated by QT whose exclusivity graphs are precisely odd cycles (Result~3). This result is not new \cite{CSW10, LSW11, AQBTC12}. The interesting result is that we have shown that there is another family of NC inequalities violated by QT whose exclusivity graphs are the complements of odd cycles (Result~4). We have described how to reach the maximum quantum value for each member of this family.

Finally, we have shown evidences that suggest that the maximum quantum violation of the inequalities in Results 3 and 4 are singled out by the E principle. This adds new examples to the list of inequalities whose maximum quantum value is singled out by this principle \cite{Cabello12b}. The fact that the exclusivity graphs of these NC inequalities are present in the exclusivity graph of {\em any} NC inequality violated by QT (Result~1) suggests that the E principle may be fundamental for quantum correlations.


\begin{acknowledgments}
We thank Costantino Budroni for useful comments. This work was supported by the Project No.\ FIS2011-29400 (MINECO, Spain).
\end{acknowledgments}


\appendix


\section{Basic exclusivity graphs inside the exclusivity graphs of some NC inequalities}
\label{App}


Table \ref{Table1} shows the number of induced basic exclusivity graphs inside some NC inequalities and Kochen-Specker (KS) proofs. The absence of induced odd antiholes may explain why the NC inequalities associated to the exclusivity graphs of type (ii) were not pointed out before. To our knowledge, the first time a type (ii) graph with $n \geq 7$ was identified as a QCG is in \cite{Web}.


\begin{table*}[htb]
\caption{\label{Table1}Number of induced basic exclusivity graphs in some NC inequalities and KS proofs. The column ``Graph'' gives the standard name in graph theory, ``Vertices'' indicates its number of vertices, ``Dimension'' indicates the minimum dimension of the quantum system needed to define events with the corresponding exclusivity relationships.}
\begin{ruledtabular}
\begin{tabular}{lcccccccc}
 NC inequality/KS proof & Graph & Vertices & Dimension & $C_5$ & $C_7$ & $\bar{C}_7$ & $C_9$ & $\bar{C}_9$ \\
 \hline
 KCBS \cite{KCBS08} & $C_5$ & 5 & 3 & 1 & 0 & 0 & 0 & 0 \\
 CHSH \cite{CHSH69} & $Ci_8 (1,4)$ & 8 & 4 & 8 & 0 & 0 & 0 & 0 \\
 $S_3$ \cite{NDSC12,ADLPBC12} & & 10 & 4 & 10 & 0 & 0 & 0 & 0 \\
 KCBS-twin \cite{Cabello12} & $J(5,2)$ & 10 & 6 & 12 & 0 & 0 & 0 & 0 \\
 Mermin \cite{Mermin90} & Complement of Shrikhande & 16 & 8 & 96 & 0 & 0 & 0 & 0 \\
 KS-18 \cite{CEG96,DHANBSC12} & & 18 & 4 & 144 & 108 & 0 & 12 & 0 \\
 YO \cite{YO12} and its tight version \cite{KBLGC12} & & 22 & 3 & 288 & 384 & 0 & 0 & 0 \\
 KS-24 \cite{Peres91} & & 24 & 4 & 576 & 576 & 0 & 192 & 0 \\
 KS-31 \cite{Peres95} & & 31 & 3 & 70 & 184 & 0 & 248 & 0 \\
 KS-33 \cite{Peres91} & & 33 & 3 & 72 & 84 & 0 & 128 & 0 \\
\end{tabular}
\end{ruledtabular}
\end{table*}



\end{document}